\documentclass[12pt,a4paper,twoside]{article}
\usepackage[T2A]{fontenc}
\usepackage{epsfig,graphicx,floatflt,amssymb,amsmath,mathrsfs,longtable,hhline,afterpage,wrapfig}
\def \insnew#1#2#3#4#5#6 {
 \begin{figure}[h]
\noindent\hspace{#4\textwidth}\resizebox{#5\textwidth}{#6\textwidth}{%
\includegraphics{#2}}
    \caption{#3}                                    
    \label{#1}                                      
  \end{figure}}                                     

\def \inspref#1#2#3#4#5 {
 \begin{figure}[h]
\noindent\hspace{#3\textwidth}\resizebox{#4\textwidth}{#5\textwidth}{%
\includegraphics{#2}}
    \label{#1}                                      
  \end{figure}}                                     

\begin{document}
\setlength{\LTcapwidth}{6in}

\renewcommand{\tablename}{\textbf{Table}}
\renewcommand{\figurename}{\textbf{Figure}}
\renewcommand{\topfraction}{1.0}
\renewcommand{\bottomfraction}{1.0}
\renewcommand{\textfraction}{0.1}
\renewcommand{\textfloatsep}{0.3cm}

\thispagestyle{empty}
\inspref{f111}{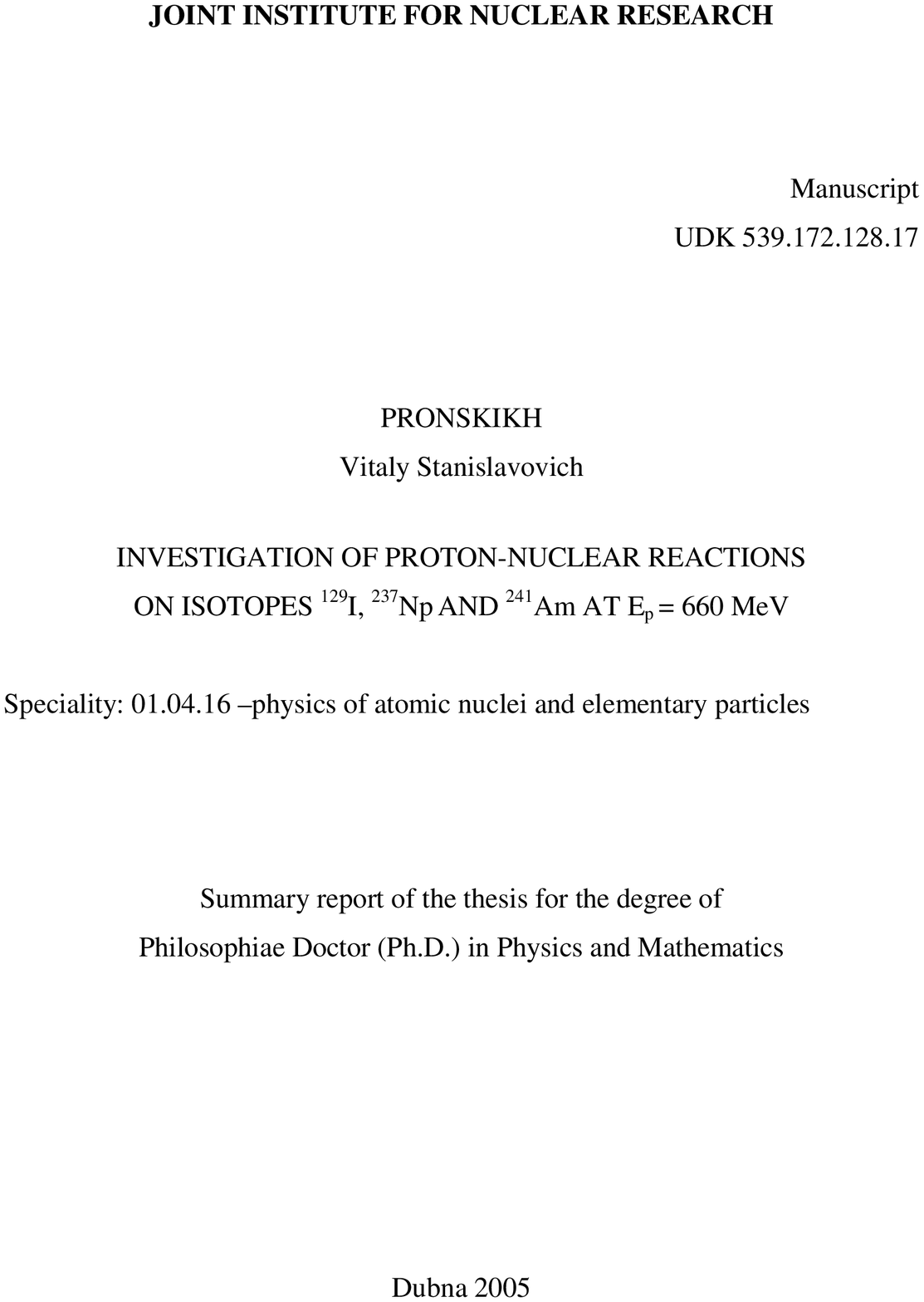}{0.0}{1.0}{1.5}
\clearpage
\newpage
\thispagestyle{empty}
\inspref{f222}{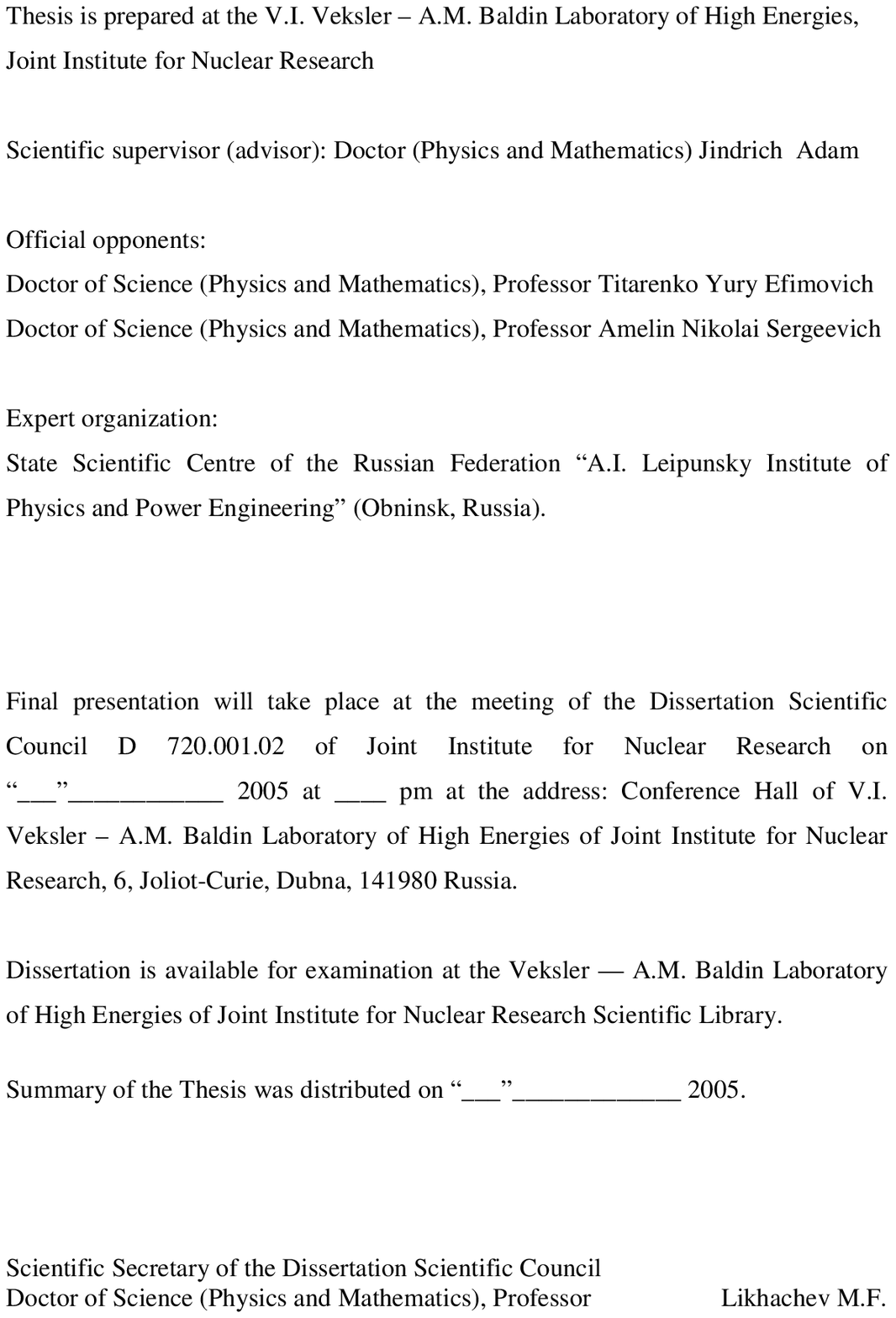}{0.0}{1.2}{1.7}
\clearpage
\newpage

\setcounter{page}{1}
\noindent\textbf{General description}: Investigations of the proton-nuclear reaction mechanisms
at intermediate and high energies became very important in recent years due to a number of reasons.
The reactions at low (<20 MeV) energies were studied during the last ten years in detail, and
the data on these reactions have been published in the form of evaluated libraries. A large amount
of data, both experimental and theoretical, in the region up to 200 MeV have recently been
collected, while the data on reaction cross sections at energies higher than 200 MeV
are incomplete and fragmentary. In this energy region changes in interaction mechanisms
became evident, spallation, fission, and fragmentation reactions start to prevail,
and the experimental methods of their investigation change accordingly. The number of
open reaction channels becomes too large for creation of the evaluated libraries, and
the calculation of necessary cross sections using theoretical computer models, which can
than be incorporated in the codes for modelling of the industrial-scale systems, becomes
more practical. Experimental investigation of proton-induced reactions, therefore, appears
to be one of the basic methods for evolving of such theories.

Among contemporary and important areas of fundamental studies that require development
of adequate nuclear reaction models are radioactive ion beams, astrophysics, and spallation
neutron sources. Spallation neutron sources consist of an intense proton source and a heavy metal
target. The fast neutrons, generated in spallation, are than moderated within heavy- or lightwater
moderators. One of the very important applications of a spallation source is Accelerator
Driven Transmutation System (hybrid reactor). In such systems spallation neutrons are used to sustain
nuclear chain reaction in the subcritical reactor, whereas the long-lived reaction fragments
can simultaneously transmute into the short-lived and stable isotopes in the subcritical blanket.
Such system can produce electrical energy, part of which is spent to supply the accelerator work.

Accumulation of large amounts of long-lived radioactive isotopes is one of the most significant
shortcomings of the present-day nuclear industry, which nevertheless remains an unavoidable
mean of energy production within the next few decades. Radiation hazard assessments show
that after isolation of the U-Pu series actinides, and such fission products as
$^{99}$Tc, $^{126}$Sn, $^{90}$Sr, $^{137}$Cs, $^{129}$I and $^{79}$Se removal,
$^{241}$Am and $^{237}$Np
remain the key sources of the influence over the population. Among actinides, the largest
contribution to the radioactivity is made by $^{241}$Am, while $^{237}$Np is predominant in
the spent fuel in mass. Furthermore, $^{237}$Np possesses a higher mobility in biosphere in comparison
with the other actinides, followed by higher probability of its penetration in body through food chains.
Assessments reveal that after transmutation of all transuranic isotopes to stable, their total
radioactivities would equalize in approximately $10^{3}$ years, rather than in $5\cdot 10^{6}$
years without such processing. Among the fission fragments, $^{129}$I seems to be one of the most
hazardous from biological standpoint.

Construction of subcritical reactor-transmuter is nowadays considered to be the most promising
method for destruction of the components of spent nuclear fuel, and in this connection the need
in data for the ADTS has grown significantly. Modelling of such systems requires experimental studying of
interaction of high-energy particles with actinide and fission product nuclei,
which can lead to advances in theoretical insights about these interactions and their computer
models. Due to that experimental formation cross sections of the residual nuclides serve the
differential parameter, comparison of which with calculations is optimal for the theoretical model
development.

\noindent\textbf{Aims of the study}: This dissertation work is aimed at both experimental
investigation of the residual nuclides formation at interactions of 660 MeV protons with
$^{129}$I, $^{237}$Np and $^{241}$Am isotopes using nuclear spectroscopy, and analysis
of the measured cross sections using modern theoretical models. For solving of this task,
the necessary spectroscopy methods were amended and developed, a program package
was created.

\noindent\textbf{Scientific novelty}: The following new results were obtained in this research~:
\begin{enumerate}
\item Techniques of the nuclide identification and short-lived $\beta$-unstable product formation
cross section determination by means of activation analysis with HPGe $\gamma$-spectrometers
have been amended using the precision nuclear spectroscopy approaches, a program package was
created.
\vskip -0.2cm
\item A technique was developed for calculation of optimal parameters of an experiment on studying
individual product nuclei formed with small cross sections and situated in complex decay chains.
\vskip -0.2cm
\item Experiments were carried out on determination of formation cross sections in reactions of
protons of 660 MeV on the isotope $^{129}$I. A total of 74 formation cross sections was determined
for this target. These data were obtained for the first time at intermediate energies.
\vskip -0.2cm
\item Experiments were carried out on determination of formation cross sections in reactions of
protons of 660 MeV on the isotope $^{237}$Np. A total of 53 formation cross sections was determined
for this target. These data were obtained for the first time at intermediate energies.
\vskip -0.2cm
\item Experiments were carried out on determination of formation cross sections in reactions of
protons of 660 MeV on the isotope $^{241}$Am. A total of 80 formation cross sections was determined
for this target. These data were obtained for the first time at high and intermediate energies.
\vskip -0.2cm
\item An extensive analysis of the data using eleven modern existing theoretical models~:
\texttt{LAHET} (Bertini+\texttt{RAL}, \texttt{ISABEL+RAL}, \texttt{INCL+RAL},
\texttt{INCL+ABLA}), \texttt{CASCADE}, \texttt{CEM95} (для $^{129}$I),
\texttt{CEM2k} (для $^{129}$I),
\texttt{CEM2k+GEM2}, \texttt{LAQGSM+GEM2}, \texttt{CEM2k+GEMINI}, \texttt{LAQGSM+GEMINI}
was performed on the basis of both qualitative and quantitative criteria, which showed
insufficient for practical purposes precision of  modelling of the reactions under study.
\end{enumerate}

\noindent\textbf{Practical value}: Results presented in the dissertation can be used in the following way:
\begin{enumerate}
\item Cross section determination methods for the short-lived nuclear reaction products can be used
in activation analysis and radiation medicine, they are used for reaction rate determination in
activation detectors and transmutation samples.
\vspace{-0.2cm}
\item Optimal experimental parameter calculation methods are used for planning, carrying out,
and data analysis of the experiments on investigation of product nuclei, formed with small cross
sections and positioned in complex decay schemes.
\vspace{-0.2cm}
\item Cross sections of the proton-induced reactions in $^{129}$I, $^{237}$Np and $^{241}$Am targets
and the results of their analysis with theoretical models are necessary for investigations of
these reactions' mechanisms as well as in applied research for development of ADS modelling tools.
\vspace{-0.2cm}
\item Obtained experimental results can be used for filling the nuclear data libraries (EXFOR, NSR).
\end{enumerate}

\noindent\textbf{Approbation of the Thesis}: The dissertation is based on the works [1-8] reported at the
seminars of the Dzhelepov Laboratory of Nuclear Problems of Joint Institute
for Nuclear Research, scientific conferences of the Association of Young Scientists and Specialists of
JINR /Dubna (1999, 2002)/, International Conferences on Nuclear Spectroscopy and Structure of Atomic Nucleus
/Obninsk (1997), Moscow (1998), Saint-Petersburg (2000), Sarov (2001), Moscow (2003)/, Conference of
the US Radiochemical Society /San-Diego (2001)/, Workshop on Nuclear Reaction Data and Nuclear Reactors
at International Centre for Theoretical Physics, Italy /Trieste (2002)/, International Conferences on Nuclear
for the Transmutation on Nuclear Waste, Germany /Darmstadt (2003)/, Specialists' Meeting Shielding Aspects
of Accelerators, Targets and Irradiated Facilities SATIF-7, Portugal /Sacavem (2004)/, International Conferences
on Nuclear Data for Science and Technology, Japan /Tsukuba (2002)/, USA /Santa Fe, NM (2004)/.

\noindent\textbf{Structure of the Thesis}: The dissertation consists of an introduction, three chapters,
conclusion, acknowledgements, and a list of references. The overall text size including 35 figures,
17 tables and a list of references containing 99 items amounts to 131 pages. The main results of the
dissertation have been published in 8 co-authored papers (67 pages of published text).
The principal contribution to these works was made by the author of this Thesis.

\noindent\textbf{Contents of the Thesis}: in \textbf{Introduction} the topicality of studies of
proton-induced reactions at high and intermediate energies is discussed, a qualitative description
of the processes of interaction of intermediate energy protons with nuclei and a survey of
the theoretical models employed is given, the aim of the study is formulated and a short summary of
the dissertation is presented.

\noindent\textbf{First Chapter} is devoted to the discussion of the problems and methods of  analysis of
$\gamma$-spectra measured with semiconductor detectors. In investigations of nuclear structure as well as
residual nuclei formation cross sections a number of tasks arise, some of which were solved in this work.
Among them are developing of the method for optimal experimental parameter calculation, precise determination
of half-lives from short-lived nuclei decays measured with semiconductor detectors, application of the precision
spectroscopy approaches in activation measurements, and automation of $\gamma$-spectra analysis.
The $\gamma$-spectra are measured in the energy range from 50 to 3500 keV with the use of the spectrometers
with HPGe-detectors. In the course of the data analysis, the radioactive background as well as the target
spectrum before irradiation were taken into consideration. In separate measurements intensities of single
and double escape peaks, which moderately depend on particular measurement geometry.
An analytical curve in of the form :
$I^{esc}_{\gamma}/I_{\gamma} = \sum_{i=1}^{4} a_i(\ln (E_\gamma))^{i}$,
fitted to the ratios of the escape peak intensities to the respective photopeaks was used for correcting
the $\gamma$-line areas in the spectra.
An 8-th degree function of the same form was also used to describe the photopeak registration efficiency
of the HPGe-detector.
In the course of the residual nuclide identification, their T$_{1/2}$ as well as their $\gamma$-transition
energies and relative intensities were taken into account, which appeared to be especially important
when analyzing complex peaks. A criterion $K=(I_{\gamma}^{lit}\cdot I_{\gamma,max}^{exp})/
(I_{\gamma}^{exp}\cdot I_{\gamma,max}^{lit})$ was used, where $I_{\gamma}^{lit}$ and $I_{\gamma,max}^{lit}$
are adopted intensities of the transition under analysis and the reference transition, while
$I_{\gamma}^{exp}$ and $I_{\gamma,max}^{exp}$  are their experimental intensities.
In case of correct identification and absence of peak overlaps, this criterion is to be equal to unity
in the range of two uncertainties; if a peak of an isotope is absent in the spectrum, $K<1$ for the
upper limit of its detection.

As the spectra investigated are very complex and the $\gamma$-line overlap probability for different isotopes
is very high, in the case of an admixture detection such lines were decomposed into components using the
least squares method. Different possibilities of genetic relations between the isotopes found were considered.
For the purpose of precision spectra analysis\footnote{Preliminary data processing was performed with
the program~\texttt{deimos32} from NPI, \v{R}e\v{z}}, spurious, background and escape peak subtraction,
their identification, decomposition and cross section calculations a package of PC programs has been created [1].

One of the most important parameters influencing reliability of radioactive nucleus identification is
its half-live $T_{1/2}$. Employing modern types of electronics units in spectrometers requires refinement
of $T_{1/2}$ measurement methods, and thorough assessment of properties of each configuration. Basically,
when a short-lived radioactive source is measured during the time comparable with its $T_{1/2}$,
deadtime of the spectrometer changes almost proportionally to its activity, and the equation for
its $\gamma$-peak area in spectra has the form~:
\begin{multline}
S_i\frac{\Delta t_i^r}{\Delta t_i^l}=
N_0\{(e^{-\lambda t_i}-e^{-\lambda(t_i+\Delta t_i^r)})+
C_1(e^{-2\lambda t_i}-e^{-2\lambda(t_i+\Delta t_i^r)})+\\
+C_2(e^{-3\lambda t_i}-e^{-3\lambda(t_i+\Delta t_i^r)})\, ,
\end{multline} $N_0$ -- initial number of nuclei,
$t_i$ -- measurement start time, $\Delta t_i^r$ и  $\Delta t_i^l$ -- real and live times of measurement.

When analyzing spectra of a $\beta^-$-decaying nuclide $^{140}$La, formed in the reaction
${}^{139}$La$(n,\gamma){}^{140}$La, measured using a $\gamma$-spectrometer assembled of a HPGe
28\%-efficiency detector \texttt{ORTEC}, a spectroscopy amplifier \texttt{CANBERRA 2026}, and
a multichannel analyzer \texttt{SPECTRUM MASTER 919} with an automatic deadtime determination,
it appeared that the deadtime found in a conventional way (using only the first term of the
equation (1)) differs from the one recommended by Nuclear Data Sheets by 12$\sigma$, whereas the
maximum deadtime of the setup amounted to 8\%.
Additional experiments were carried out, in which $^{139}$La samples were irradiated in a thermal
neutron flux of the microtron MT-25, deadtimes varying from 1 to 50\%. It was shown [2] that
at a 50\% deadtime, its underestimation by the setup leads to depreciation of $\gamma$-line areas
by 10\%, while the correct $T_{1/2}$ value is attained only by using of all the three terms in (1).
These experiments allowed to clarify the reasons of observed differences of $^{140}$La half-lives
determined in a series of activation measurements with large loads from the adopted value, and correspond
them to systematic uncertainties arising from neglecting the second and third terms in (1), as well as
to determine the value of $T_{1/2}(^{140}$La$)$ = \textbf{1.6808 (18) days} [2].

\begin{wrapfigure}[13]{i}{8.5cm}
{\includegraphics[scale=0.3]{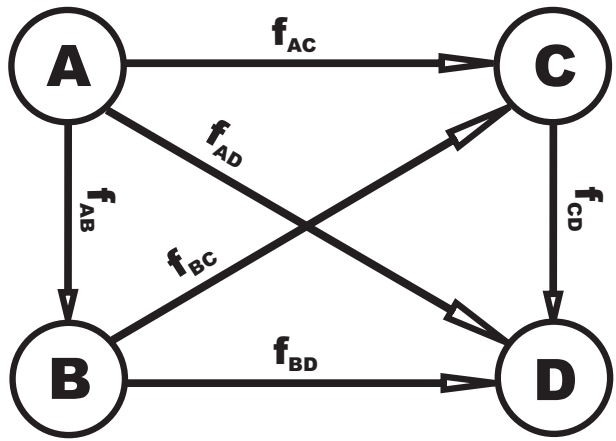}
\caption{General view of decay scheme of four nuclei
(or their excited states).}\label{figdecay}}
\end{wrapfigure}
\begin{wrapfigure}[13]{r}{7.7cm}
{\includegraphics[scale=0.3]{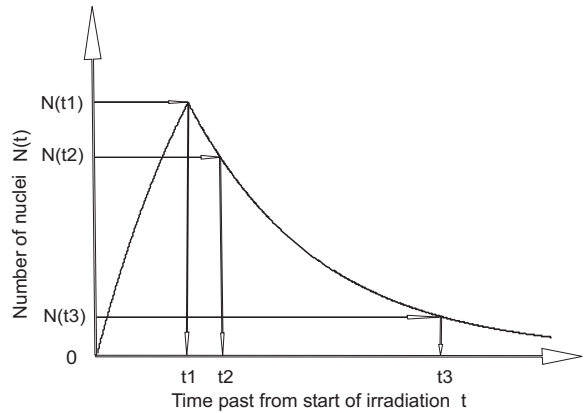}
\caption{Nuclei accumulation diagram in a decay chain.}\label{rist3}}
\end{wrapfigure}

~~~~In experiments with spallation products it often appears that the product nuclei formed find themselves
in complex decay chains, and several targets of the same isotope are irradiated and measured several times,
irradiation time $t_1$, delay time $t_2$ and measurement time $t_3$ varying in a broad range. In this connection
two problems were solved in this work~: to determine formation cross sections of daughter nuclei in a
complex chain, and to find optimum $t_1$,  $t_2$  and  $t_3$ for accumulation of desired nuclide situated
in such a chain. For the practical purpose it is enough to confine ourself to a chain of four nuclei
(see \textbf{Figure \ref{figdecay}}), which limitation is required by finite statistics in $\gamma$-spectra.
The number of genetically related nuclei $N_A(t)$, $N_B(t)$, $N_C(t)$ and $N_D(t)$ during the irradiation,
delay, or measurement is described by the simultaneous differential equations~:
\begin{eqnarray*}
\left\{
\begin{array}{lll}
dN_{A}(t)/dt&=&Q_{A}-\lambda_{A}N_{A}(t)\\
dN_{B}(t)/dt&=&Q_{B}-\lambda_{B}N_{B}(t)+
f_{AB}\lambda_{A}N_{A}(t)\\
dN_{C}(t)/dt&=&Q_{C}-\lambda_{C}N_{C}(t)+
f_{AC}\lambda_{A}N_{A}(t)+
f_{BC}\lambda_{B}N_{B}(t)\\
dN_{D}(t)/dt&=&Q_{D}-\lambda_{D}N_{D}(t)+
f_{AD}\lambda_{A}N_{A}(t)+
f_{BD}\lambda_{B}N_{B}(t)+f_{CD}\lambda_{C}N_{C}(t)\, ,
\end{array}
\right.
\end{eqnarray*}
where $Q$ is formation rate of a given isotope (during delay and measurement $Q = 0$). Such sets were
solved for each of the $t_1$, $t_2$ and $t_3$ intervals (see \textbf{Figure~\ref{rist3}}), while their
solution appeared to be convenient to find in the form of recurrence relations.
Thus, for estimation of independent formation
cross sections, the following set of linear equations is solved by the least squares method~:
\begin{equation*}
\left\{
\begin{array}{lll}
S_{\gamma}^i\frac{\Delta t_3^{r,i}}{\Delta t_3^{l,i}}K_{x_A}
&=&A_A^i\sigma_A\,,\\
S_{\gamma}^i\frac{\Delta t_3^{r,i}}{\Delta t_3^{l,i}}K_{x_B}
&=&A_B^i\sigma_A + B_B^i\sigma_B \,,\\
S_{\gamma}^i\frac{\Delta t_3^{r,i}}{\Delta t_3^{l,i}}K_{x_C}
&=&A_C^i\sigma_A + B_C^i\sigma_B + C_C^i\sigma_C \,,\\
S_{\gamma}^i\frac{\Delta t_3^{r,i}}{\Delta t_3^{l,i}}K_{x_D}
&=&A_D^i\sigma_A + B_D^i\sigma_B + C_D^i\sigma_C + D_D^i\sigma_D\,,\\
\end{array}
\right.
\end{equation*}
where $S_{\gamma}^{i}$ is the peak area with $E_{\gamma}$ in $i$-th measurement,
$K_{xj}$ is a function depending on the $\gamma$-detector efficiency, $\gamma$-quanta intensity,
number of target nuclei, and number of protons, $A_j$, $B_j$, $C_j$ and $D_j$ are some recurring
equations deduced. Analogous formulae were obtained also for cumulative cross sections. All the recurring
coefficients are presented in [3] in detail.
~~As the main criterion of success in an experiment is the measurement of an effect (peak area) with
the least possible uncertainty, the peak area $N$, referring to decay of $j$-th nuclei (for instance,
of sort $B$), is proportional to the number of nuclei of this sort $\Re_j(t_3)$, decayed during the
measurement time, $N=k\Re_j(t_3)$, where the forms of $\Re_j$ for various nuclei of the chain are
deduced in [3]. The other nuclei ($A$, $C$, $D$) produce a spurious background
$\Re_f(t_3)=\Re_A(t_3)+\Re_C(t_3)+\Re_D(t_3)$, thus the number of collected background events
$\Phi=k\Re_f(t_3)$, and the total number of events $T=N+\Phi$.

On assumption of cyclic measurement regime, for the number of cycles one can write~:
$c=\Delta t_T/(\Delta t_1+\Delta t_2+\Delta t_3)$. Taking into account that the decays of $j$ and $f$
nuclei are statistically independent, and the uncertainties of numbers of collected events are added in
the following way~:
$\Delta^2_N=\Delta^2_T+\Delta^2_\Phi$, and the relative measurement error equals to $\Delta N/N$,
the following quality criterion of an experiment (figure of merit) was obtained~:
\begin{eqnarray*}
K_{\mbox{\small e}}=\frac{\Re_j(t_3)}{\sqrt{\Re_j(t_3)+2\Re_f(t_3)}}\sqrt{c k}\, .
\end{eqnarray*}
~~~~As at the planning stage of an experiment $k$ and $t_T$ are often to be defined in advance,
a provision made in the program \texttt{optimum}, created in this work, to vary each of the parameters
$t_1$,  $t_2$  and  $t_3$ with a predefined step, finding maxima of the objective function
$K_{\mbox{\small e}}$. The calculations performed for the isobar with $A = 152$ were published in [3].
The approach is proposed to use for selection of optimal conditions for measurements of spallation
products, especially those formed with small cross sections, and if the problem to solve is to study individual
isotopes. This method will give the best results in the case of using, for example, a device for
transporting products to detectors.

\begin{wrapfigure}[15]{i}{8.2cm}
{\includegraphics[scale=0.42]{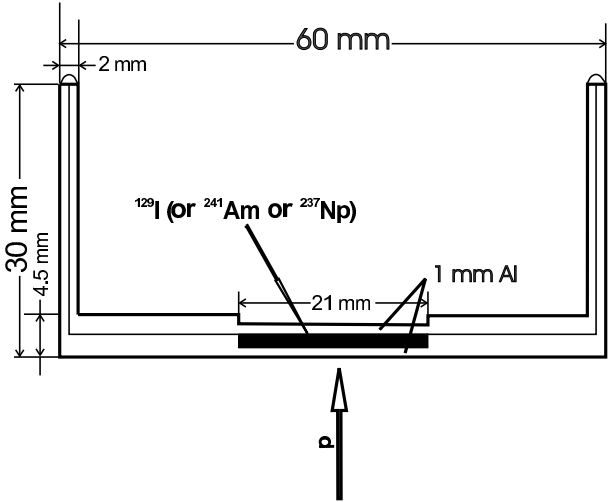}
\caption{Weld sealed aluminium containers with radioactive samples used, prepared in IPPE,
Obninsk.}\label{figtarg}}
\end{wrapfigure}

\noindent In the \textbf{Chapter Two} experiments on determination of proton-induced reaction cross sections
on isotopes $^{129}$I, $^{237}$Np and $^{241}$Am at the proton energy 660 MeV are described. The experiments
were carried out at the extracted proton beam of Phasotron with the beam current 1.2 mkA. The targets
consisted of NpO$_{2}$, AmO$_{2}$ and NaI (15\% $^{127}$I + 85\% $^{129}$I). The samples were
weld-sealed in aluminium containers, 79 g in weight. (see \textbf{Figure~\ref{figtarg}}).
Characteristics of the targets and beam are presented in \textbf{Table~\ref{table1}}.
\begin{table}[t]
\begin{center}
\caption{Characteristics of the $^{237}$Np, $^{241}$Am and $^{129}$I targets and the proton beam.}\label{table1}
\vskip 3mm
\begin{tabular}{|l|c|c|c|c|c|c|}  \hline
Target nuclide   &   \multicolumn{2}{|c|}{$^{237}$Np$_{93}$}&
               \multicolumn{2}{|c|}{$^{241}$Am$_{95}$}&
               \multicolumn{2}{|c|}{$^{129}$I$_{53}$}\\ \hline
Half-live, y   &    \multicolumn{2}{|c|}{ 2.144(7)$10^6$}   &
\multicolumn{2}{|c|}{432.2(7)} &\multicolumn{2}{|c|}{15.7(4)$10^6$} \\ \hline
Target weight, g            &  0.742&   0.742  &  0.177 &  0.183 & 0.500 & 0.500 \\
\hline Target thickness, mm        &  0.193&  0.193  &  0.043 &  0.044 & 0.395& 0.395\\
\hline Initial activity, mCi   &0.523&  0.523  &     601 &    621  & 0.063& 0.063 \\
\hline Beam intensity, 10$^{14}$ p/min & 2.64 &  2.66  & 2.72 &   2.58 & 2.68& 2.68\\
\hline Irradiation time, min    &   5  &    30 & 5  &    30  & 5& 30\\ \hline
\end{tabular}
\end{center}
\end{table}
Two targets of each nuclide were irradiated in all. The beam size and position were checked by a
two-coordinate proportional chamber. For the purpose of monitoring, reaction $^{27}$Al$(p, 3pn)^{24}$Na
was employed, and Al foils of the same size as the target nuclide spots and 99 mg in weight were
used as monitors.
~~For reducing of the targets' own background, a shielding of a 10 mm thick Pb, 2 mm thick Cd,
and 1 mm thick Cu plates was mounted on the detector. Activity induced in the targets was counted with
the help of three semiconductor $\gamma$-detectors~: 1) the $^{129}$I target at a HPGe detector with
efficiency $\varepsilon_\gamma$ = 50\% and energy resolution $E_\gamma$ = 2.15 keV at the line 1332 keV
($^{60}$Co); 2) the $^{241}$Am target at a HPGe detector with $\varepsilon_\gamma$ = 20\% and
$E_\gamma$ = 1.8 keV; 3) the $^{237}$Np at a Ge(Li) detector with  $\varepsilon_\gamma$ = 4.8\% and
$E_\gamma$ = 2.6 keV. Acquiring of $\gamma$-spectra was carried out with the help of \texttt{MASTER 921}
($^{241}$Am and $^{129}$I) and \texttt{MASTER 919} ($^{237}$Np) analyzers, which automatically determined
deadtime of the spectrometer.

Counting of the first set of targets was started 10 minutes after the end of their irradiation, 17 spectra
were taken from each sample, measurement time ranging from 5 minutes to 3 hours, initial distances from
samples to respective detectors made 225 cm for $^{129}$I, 150 cm for $^{241}$Am and 100 cm for $^{237}$Np.
Measurement of the second set of $^{129}$I samples (30-minute exposures) was started 20 hours after the end of irradiation
(17 hours for $^{241}$Am and $^{237}$Np), 13 spectra were taken during 33 days (11 during 30 days for
$^{241}$Am and
$^{237}$Np), counting time varied from 5 to 66 hours (50 hours for $^{241}$Am and $^{237}$Np).
Using the developed program package [2], energies and intensities of $\gamma$-transitions in the residual
nuclei were determined, as well as the peak registration limits on a given background.
With the use of the programs, background, single and double
escape peaks were found, and if their overlaps with the photopeaks were detected, intensities of the latter
were corrected.
Using corrected $\gamma$-lines, half-lives of isotopes were determined where possible. Residuals
were than identified using energies of their $\gamma$-lines, half-lives, and if several $\gamma$-lines
of the same nucleus were found, also ratios of their intensities, which was compared with that
from recommended data sets. In total, more than 2800 $\gamma$-lines in the spectra of $^{129}$I target,
more than 1000 in the spectra of $^{241}$Am, and more than 500 in the spectra of $^{237}$Np were found,
formation cross sections of 74 product nuclei in $^{129}$I, 80 -- in $^{241}$Am and 53 -- in $^{237}$Np targets
were determined.
Final experimental results for all the three targets are presented in
\textbf{Tables~\ref{tablei}-\ref{tableam}}. In the \textbf{Tables}, cumulative cross sections are denoted by
the letter ``C'', independent -- ``I'', isomer transition -- ``IT'', $\beta^+$, $\beta^-$ -- beta-decay,
$\varepsilon$ -- electron capture.
{\small
\begin{longtable}{|r|c|c|c||r|c|c|c|}
\caption{Experimental data for the target $^{129}$I$+{}^{127}$I.}\label{tablei}\\
\hline
  Isotope   & $T_{1/2}$ &   Decay           & $\sigma_{exp}$,&  Isotope & $T_{1/2}$ &   Decay           & $\sigma_{exp}$,\\
           &           &    mode            &    mb          &          &           &   mode            &    mb\\\hline
\endfirsthead
\caption{Experimental data for the target $^{129}$I$+{}^{127}$I (continued).}\\
\hline
  Isotope   & $T_{1/2}$ &   Decay         & $\sigma_{exp}$,&  Isotope  & $T_{1/2}$ &  Decay           & $\sigma_{exp}$,\\
           &           &    mode          &    mb          &          &           &   mode       &    mb\\\hline
\endhead
\hline
\endfoot
$^{44m}$Sc   &       2.44 d  &      I(IT,$\varepsilon$)  &      0.20(4)  &   $^{104}$Ag   &      69.2 m   &      I($\varepsilon$, $\beta^+$)     &   8.3(8)  \\ \hline
$^{46}$Sc    &     83.83 d   &       C($\beta^-$)   &      0.36(4)  &   $^{105}$Ag   &    41.29 d    &      C($\varepsilon$)         &     14.3(17)  \\ \hline
$^{48}$V     &      15.97 d  &      C($\beta^+$,$\varepsilon$)      &      0.58(6)  &   $^{106}$Ag   &       8.46 d  &      I($\varepsilon$)         &   7.5(7)  \\ \hline
$^{52}$Mn    &       5.29 d  &      C($\varepsilon$, $\beta^+$)     &      0.37(5)  &   $^{108}$In   &         58 m  &      I($\beta^+$,$\varepsilon$)      &   7.1(7)  \\ \hline
$^{56}$Co    &      78.8  d  &      C($\varepsilon$, $\beta^+$)     &      0.11(4)  &   $^{109}$In   &       4.20 h  &      I($\varepsilon$, $\beta^+$)     &     12.1(12)  \\ \hline
$^{58}$Co    &     70.92 d   &      C($\varepsilon$, $\beta^+$)     &      0.47(15) &   $^{109}$Sn   &     18.0  m   &      C($\varepsilon$, $\beta^+$)     &     3.02(30)  \\ \hline
$^{59}$Fe    &     44.5  d   &      C($\beta^-$)        &      0.065(7) &   $^{110m}$Ag  &    249.9  d   &      I($\beta^-$,IT)     &     1.50(18)  \\ \hline
$^{65}$Zn    &    244.1  d   &      C($\varepsilon$, $\beta^+$)     &      0.88(9)  &   $^{110}$In   &       4.9  h  &      I($\varepsilon$)         &     11.7(12)  \\ \hline
$^{72}$As    &     26 h      &      I($\beta^+$,$\varepsilon$)      &      0.93(9)  &   $^{110m}$In  &     69.1  m   &      C($\varepsilon$)         &   6.4(8)  \\ \hline
$^{72}$Se    &      8.4  d   &      C($\varepsilon$)             &      0.59(15) &   $^{113}$Sn   &    115.1  d   &      C($\varepsilon$)         &     27.2(30)  \\ \hline
$^{74}$As    &     17.78 d   &      I($\varepsilon$, $\beta^{+/-}$)  &      0.85(9)  &   $^{114m}$In  &     49.51 d   &      I(IT, $\varepsilon$)     &   7.1(7)  \\ \hline
$^{76}$Br    &      16.2  h  &      C($\beta^+$,$\varepsilon$)      &      0.77(9)  &   $^{114}$Sb   &     3.49  m   &      C($\varepsilon$, $\beta^+$)     &     2.92(35)  \\ \hline
$^{77}$Br    &       2.38 d  &      C($\varepsilon$, $\beta^+$)     &      0.64(15) &   $^{115m}$In  &       4.49 d  &      I(IT, $\beta^-$)    &     15.6(35)  \\ \hline
$^{83}$Rb    &      86.2  d  &      C($\varepsilon$)             &      0.40(8)  &  $^{115}$Sb   &     32.1  m   &      C($\varepsilon$, $\beta^+$)     &     20.0(30)  \\ \hline
$^{84}$Rb    &     32.87 d   &      I($\varepsilon$, $\beta^{+/-}$)  &      0.12(4)  &   $^{116m}$In  &    54.15 m    &      I($\beta^-$)        &     2.72(42)  \\ \hline
$^{85}$Sr    &     64.84 d   &      C($\varepsilon$)             &      1.49(16) &   $^{116}$Sb   &     15.8  m   &      I($\beta^+$,$\varepsilon$)      &   2.0(4)  \\ \hline
$^{86}$Y &     14.74 h   &      C($\beta^+$,$\varepsilon$)      &      0.69(25) &   $^{116m}$Sb  &     60.3  m   &      I($\varepsilon$, $\beta^+$)     &     11.6(14)  \\ \hline
$^{87}$Y &       3.35 d  &      C($\varepsilon$, $\beta^+$)     &      1.15(11) &   $^{116}$Te   &       2.49 h  &      C($\varepsilon$, $\beta^+$)     &    9.9(10)    \\ \hline
$^{88}$Y &    106.6  d   &   I($\varepsilon$, $\beta^+$)    &      0.36(10) &   $^{117}$In   &      43.8 m   &      I($\beta^-$)        &   1.6(3)  \\ \hline
$^{88}$Zr   &      83.4  d  &      C($\varepsilon$)             &      1.4(4)   &   $^{117}$Te   &       1.03 h  &      C($\varepsilon$, $\beta^+$)     &     15.4(15)  \\ \hline
$^{89}$Zr   &       3.27 d  &      C($\varepsilon$, $\beta^+$)     &      1.51(15) &   $^{118m}$Sb  &       5.00 h  &      I($\varepsilon$, $\beta^+$)     &     11.1(12)  \\ \hline
$^{90}$Nb   &    14.60 h    &      C($\beta^+$,$\varepsilon$)      &      1.18(13) &   $^{118}$Te   &       6.00 d  &      I($\varepsilon$)             &     13.9(14)  \\ \hline
$^{92m}$Nb   &    10.15 d    &      I($\varepsilon$, $\beta^+$)     &      0.11(4)  &   $^{118}$I    &      13.7 m   &      C($\beta^+$,$\varepsilon$)      &   3.3(4)  \\ \hline
$^{93m}$Mo   &       6.85 h  &      I(IT, $\varepsilon$)     &      0.94(30) &   $^{119}$Te   &     16.05 h   &      C($\varepsilon$, $\beta^+$)     &     11.5(12)  \\ \hline
$^{93}$Tc   &       2.75 h  &      C($\varepsilon$, $\beta^+$)     &      2.14(24) &   $^{119m}$Te  &       4.69 d  &      I($\varepsilon$, $\beta^+$)     &     16.1(15)  \\ \hline
$^{94}$Tc   &       4.88 h  &      I($\varepsilon$, $\beta^+$)     &      1.79(17) &   $^{120m}$Sb  &       5.76 d  &      I($\varepsilon$)             &   6.3(6)  \\ \hline
$^{94m}$Tc   &        52  m  &      I($\beta^+$,$\varepsilon$)      &      0.44(8)  &   $^{120}$I    &       1.35 h  &      C($\beta^+$,$\varepsilon$)      &     10.2(12)  \\ \hline
$^{95}$Nb   &     34.98 d   &      C($\beta^-$)        &      0.36(5)  &   $^{120m}$I   &      53.0 m   &      I($\beta^+$,$\varepsilon$)      &   2.9(3)  \\ \hline
$^{95}$Tc   &      20.0  h  &      C($\varepsilon$)         &      3.00(34) &   $^{121}$Te   &      16.8  d  &      C($\varepsilon$)         &     17.9(18)  \\ \hline
$^{96}$Tc   &       4.28 d  &      I($\varepsilon$)         &      2.50(25) &   $^{121m}$Te  &     154.0 d   &      I(IT, $\varepsilon$)     &     13.3(16)  \\ \hline
$^{99}$Rh   &      16.0  d  &   I($\varepsilon$, $\beta^+$)    &      0.84(25) &   $^{122}$Sb   &       2.70 d  &   C($\varepsilon$,$\beta^{+/-}$)  &   7.7(9)  \\ \hline
$^{100}$Rh   &      20.8  h  &      I($\varepsilon$, $\beta^+$)     &      4.21(50) &   $^{123}$I    &      13.2  h  &      C($\varepsilon$)         &     25.0(24)  \\ \hline
$^{100}$Pd   &       3.63 d  &      C($\varepsilon$)         &      2.85(33) &   $^{124}$I    &       4.18 d  &      I($\varepsilon$, $\beta^+$)     &     26.3(30)  \\ \hline
$^{101m}$Rh  &      4.34 d   &      I($\varepsilon$, $\beta^+$)     &      2.9(9)   &   $^{126}$Sb   &      12.4  d  &      I($\beta^-$)        &     0.54(5)   \\ \hline
$^{101}$Pd   &       8.47 h  &      C($\varepsilon$, $\beta^+$)     &      6.3(8)   &   $^{126}$I    &     13.02 d   &      I(IT)        &   35(5)   \\ \hline
$^{102}$Rh   &       2.9  Y  &      I($\varepsilon$)         &      2.98(30) &   $^{127}$Xe   &     36.46 d   &      C($\varepsilon$)         &   3.8(4)  \\ \hline
$^{103}$Ru   &     39.25 d   &      C($\beta^-$)    &      0.43(5)  &   $^{128}$I    &     25.0  m   &      I($\beta^-$,$\varepsilon$)  &   31(5)   \\ \hline
\end{longtable}
}
{\small
\begin{longtable}{|r|c|c|c||r|c|c|c|}
\caption{Experimental data for the target ${}^{237}$Np.}\label{tablenp}\\
\hline
Isotope & $T_{1/2}$& Decay  & $\sigma_{exp}$, & Isotope & $T_{1/2}$& Decay    & $\sigma_{exp}$,\\
         &          & mode & mb        &                &          & mode & mb\\\hline
       \endfirsthead
\caption{Experimental data for the target ${}^{237}$Np (continued).}\\
\hline
Isotope & $T_{1/2}$& Decay  & $\sigma_{exp}$, & Isotope & $T_{1/2}$& Decay    & $\sigma_{exp}$,\\
         &          & mode & mb        &                &          & mode & mb\\\hline
\endhead
\hline
\endfoot
$^{48}$Sc   &   1.82    d   &   I($\beta^{-}$)  &   5.2(15)         &   $^{122}$Sb  &   2.7 d       &   C($\beta^{-}$)  &   19(4)    \\\hline
$^{48}$V    &   15.97   d   &   C($\beta^{+}$)  &   0.89(20)        &   $^{124}$Sb  &   60.2    d   &   C($\beta^{-}$)  &   16.7(20)    \\\hline
$^{56}$Mn   &   2.58    h   &   C($\beta^{-}$)  &   25.4(45)        &   $^{126}$Sb  &   12.4    d   &   C($\beta^{-}$)  &   13.9(20)    \\\hline
$^{74}$As   &   17.77   d   &   I($\beta^{-}$,$\varepsilon$)&   3.6(5)         &   $^{127}$Sb  &   3.85    d   &   C($\beta^{-}$)  &   14.7(20)    \\\hline
$^{83}$Rb   &   86.2    d   &   C($\varepsilon$)           &   6.1(8)          &   $^{128}$Sb  &   9.01    h   &   C($\beta^{-}$)  &   90(10)    \\\hline
$^{84}$Rb   &   32.77   d   &   C($\beta^{+}$,$\beta^{-}$)&   13.3(16)       &   $^{132}$Te  &   3.26    d   &   C($\beta^{-}$)  &   13(3)   \\\hline
$^{86}$Rb   &   18.63   d   &   C($\beta^{-}$)   &   17.7(20)    &   $^{133m}$Te &   55.4    m   &   I($\beta^{+}$,IT)   &   18(4)    \\\hline
$^{85}$Sr   &   64.84   d   &   C($\varepsilon$)           &   9.6(20)         &   $^{124}$I   &   4.18    d   &   I($\beta^{+}$,$\varepsilon$)  &   17.3(20)    \\\hline
$^{91}$Sr   &   9.63    h   &   C($\beta^{-}$)  &   29(3)           &   $^{131}$I   &   8.04    d   &   C($\beta^{-}$)  &   20(4)    \\\hline
$^{87}$Y    &   3.35    d   &   C($\beta^{+}$, $\varepsilon$)           &   6.7(8)          &   $^{134}$I   &   52.5    m   &   C($\beta^{-}$)  &   12.9(15)    \\\hline
$^{88}$Y    &   106.6   d   &   C($\beta^{+}$, $\varepsilon$)           &   10.4(19)        &   $^{136}$Cs  &   13.16   d   &   C($\beta^{-}$)  &   9.1(13)  \\\hline
$^{89}$Zr   &   3.27    d   &   C($\varepsilon$)           &   4.6(5)          &   $^{138}$Cs  &   33.41   m   &   C($\beta^{-}$)  &   14.9(29)    \\\hline
$^{95}$Zr   &   64.02   d   &   C($\beta^{-}$)  &   59(6)           &   $^{131}$Ba  &   11.5    d   &   C($\beta^{+}$,$\varepsilon$)   &   71(13)  \\\hline
$^{95}$Nb   &   33.15   d   &   C($\beta^{-}$)  &   22(4)           &   $^{140}$Ba  &   12.75   d   &   C($\beta^{-}$)  &   23(4)    \\\hline
$^{99}$Mo   &   2.75    d   &   C($\beta^{-}$)  &   73(13)          &   $^{145}$Eu  &   5.93    d   &   C($\beta^{+}$,$\varepsilon$)   &   0.83(9)   \\\hline
$^{95m}$Tc  &   61  d       &   C($\beta^{+}$,$\varepsilon$)&   2.3(4)         &   $^{146}$Eu  &   4.59    d   &   I($\beta^{+}$,$\varepsilon$)   &   4.2(6)    \\\hline
$^{96}$Tc   &   4.98 d      &   C($\beta^{+}$,$\varepsilon$)          &   5.7(9)          &   $^{147}$Eu  &   24.1    d   &   C($\beta^{+}$,$\varepsilon$)   &   1.9(6)  \\\hline
$^{103}$Ru  &   39.26   d   &   C($\beta^{-}$)  &   63(7)           &   $^{146}$Gd  &   48.27   d   &   C($\varepsilon$)   &   1.39(16)   \\\hline
$^{105}$Ru  &   4.44    h   &   C($\beta^{-}$)  &   19.6(20)        &   $^{152}$Tb  &   17.5    h   &   C($\beta^{+}$,$\varepsilon$)   &   27(3)   \\\hline
$^{106m}$Rh &   2.17    h   &   I($\beta^{-}$)  &   55(9)           &   $^{171}$Lu  &   8.24    d   &   C($\beta^{+}$,$\varepsilon$)   &   2.4(10) \\\hline
$^{106m}$Ag &   8.28    d   &   I($\beta^{+}$,$\varepsilon$)   &   6.2(8)      &   $^{185}$Os  &   93.6    d   &   C($\varepsilon$)   &   2.8(4)    \\\hline
$^{110m}$Ag &   249.49  d   &   I($\beta^{-}$,$\varepsilon$)  &   18.0(20)        &   $^{188}$Pt  &   10.2    d   &   C($\varepsilon$)   &   0.46(8) \\\hline
$^{115}$Cd  &   2.23    d   &   C($\beta^{-}$)  &   65(12)          &   $^{206}$Po  &   8.8 d   &   C($\beta^{+}$,$\varepsilon$)   &   3.8(7)  \\\hline
$^{117m}$Cd &   3.46    h   &   C($\beta^{-}$)  &   17(4)           &   $^{230}$Pa  &   17.4    d   &   I($\beta^{+}$,$\varepsilon$)   &   1.6(3)    \\\hline
$^{125}$Sn  &   9.64    d   &   C($\beta^{-}$)   &   6.6(11)        &   $^{234}$Np  &   4.4 d   &   C($\beta^{+}$,$\varepsilon$)  &   2.2(4)  \\\hline
$^{118m}$Sb &   5   h       &   I($\beta^{+}$,$\varepsilon$)   &   10.4(13)    &   $^{238}$Np  &   2.12    d   &   I($\beta^{-}$)  &   16(3)    \\\hline
$^{120m}$Sb  &   5.76   d   &   I($\beta^{+}$,$\varepsilon$)   &   14.7(16)   &       &           &       &       \\\hline
\end{longtable}
}
{\small
\begin{longtable}{|r|c|c|c||r|c|c|c|}
\caption{Experimental data for the target ${}^{241}$Am.}\label{tableam}\\
\hline
Isotope & $T_{1/2}$& Decay  & $\sigma_{exp}$, & Isotope & $T_{1/2}$& Decay   & $\sigma_{exp}$,\\
         &         & mode   & mb        &                &          &  mode  & mb\\\hline
  \endfirsthead
\caption{Experimental data for the target ${}^{241}$Am (continued).}\\
\hline
Isotope & $T_{1/2}$& Decay  & $\sigma_{exp}$, & Isotope & $T_{1/2}$& Decay   & $\sigma_{exp}$,\\
         &         & mode   & mb        &                &          &  mode  & mb\\\hline
\endhead
\hline
\endfoot
$^{48}$Sc   &   1.82 d &   I($\beta^-$) &   1.11(20)    &   $^{108m}$Rh &   6 m &   I($\beta^-$) &   11.6(15)    \\\hline
$^{48}$V    &   15.97 h &   C($\beta^+$) &   3.4(5)  &   $^{112}$Pd  &   21.01 h &   C($\beta^-$) &   21.0(28)    \\\hline
$^{52}$V    &   3.74 m  &   C($\beta^-$) &   2.3(6)  &   $^{106m}$Ag &   8.28 d  &   I($\beta^{+/-}$)    &   2.5(3)  \\\hline
$^{52}$Mn   &   5.59 d  &   C($\beta^+$,$\varepsilon$)  &   1.74(28)    &   $^{110m}$Ag &   249.49 d    &   I($\beta^-$,$\varepsilon$)  &   11.6(24)    \\\hline
$^{54}$Mn   &   312.3 d &   I($\varepsilon$)   &   10.1(14)    &   $^{112}$Ag  &   3.130 h &   I($\beta^-$) &   20(4)   \\\hline
$^{56}$Mn   &   2.58 h  &   C($\beta^-$) &   6.7(16) &   $^{115}$Cd  &   2.23 d &   C($\beta^-$) &   19.2(28)    \\\hline
$^{72}$Ga   &   14.1 h  &   C($\beta^-$) &   1.5(7)  &   $^{117m}$Cd &   3.46 h  &   C($\beta^-$) &   6.9(7)  \\\hline
$^{72}$As   &   26 h    &   C($\beta^+$,$\varepsilon$)  &   4.2(5)  &   $^{116m}$In &   54.41 m &   I($\beta^-$) &   16.4(24)    \\\hline
$^{76}$As   &   1.08 d  &   I($\beta^+$) &   4.5(16) &   $^{117m}$In &   116.2 m &   C($\beta^-$) &   21(3)   \\\hline
$^{76}$Br   &   16.2 h  &   C($\beta^+$) &   0.6(18) &   $^{118m}$In &   4.45 m  &   I($\beta^-$) &   6.5(9)  \\\hline
$^{82}$Br   &   35.3 h  &   I($\beta^-$) &   8.0(11) &   $^{118m}$Sb &   5 h &   I($\beta^+$,$\varepsilon$)  &   7.6(14) \\\hline
$^{84m}$Br  &   6 m &   C($\beta^-$) &   2.7(6)  &   $^{120}$Sb  &   15.89 m &   I($\beta^+$,$\varepsilon$)  &   10.8(14)    \\\hline
$^{84}$Br   &   31.8 m  &   I($\beta^-$) &   9.2(14) &   $^{122}$Sb  &   2.7 d   &   C($\beta^-$,$\varepsilon$)  &   14.0(19)    \\\hline
$^{82m}$Rb  &   6.47 h  &   C($\beta^+$,$\varepsilon$)  &   2.1(10) &   $^{124}$Sb  &   60.2 d  &   C($\beta^-$) &   10.2(14)    \\\hline
$^{84m}$Rb  &   32.77 d &   C($\beta^{+/-}$)    &   6.9(11) &   $^{126}$Sb  &   12.46 d &   C($\beta^-$) &   7.3(14) \\\hline
$^{86}$Rb   &   18.63 d &   C($\beta^-$) &   2.02(29)    &   $^{127}$Sb  &   3.85 d  &   C($\beta^-$) &   7.3(12) \\\hline
$^{89}$Rb   &   15.15 m &   C($\beta^-$) &   11.1(15)    &   $^{128}$Sb  &   9.01 h  &   C($\beta^-$) &   3.3(10) \\\hline
$^{91}$Sr   &   9.63 h  &   C($\beta^-$) &   15.0(1.7)   &   $^{119m}$Te &   4.7 d   &   I($\beta^+$,$\varepsilon$)  &   3.7(5)  \\\hline
$^{92}$Sr   &   2.71 h  &   C($\beta^-$) &   11.8 (17)   &   $^{121}$Te  &   16.78 d &   C($\beta^+$,$\varepsilon$)  &   5.3(10) \\\hline
$^{93}$Sr   &   7.42 m  &   C($\beta^-$) &   10.4(22)    &   $^{131m}$Te &   30 h    &   I($\beta^-$) &   6.5(11) \\\hline
$^{84m}$Y   &   39.5 m  &   I($\beta^+$,$\varepsilon$)  &   3.1 (9) &   $^{132}$Te  &   3.20 d  &   C($\beta^-$) &   6.7(12) \\\hline
$^{87}$Y    &   3.35 d  &   C($\beta^+$,$\varepsilon$)  &   4.4(7)  &   $^{124}$I   &   4.18 d  &   I($\beta^+$,$\varepsilon$)  &   10.6(19)    \\\hline
$^{88}$Y    &   106.65 d    &   C($\beta^+$,$\varepsilon$)  &   6.2(10) &   $^{126}$I   &   13.11 d &   I($\beta^{+/-}$)    &   6.8(18) \\\hline
$^{91m}$Y   &   49.71 m &   C($\beta^-$) &   14(4)   &   $^{130}$I   &   12.36 h &   I($\beta^-$) &   10.0(19)    \\\hline
$^{95}$Y    &   10.3 m  &   C($\beta^-$) &   17(4)   &   $^{131}$I   &   8.02 d  &   C($\beta^-$) &   14.1(23)    \\\hline
$^{89}$Zr   &   3.27 d  &   C($\beta^+$,$\varepsilon$)  &   3.8(7)  &   $^{132}$I   &   2.30 h  &   I($\beta^-$) &   8.3(12) \\\hline
$^{95}$Zr   &   64.02 d &   C($\beta^-$) &   36(5)   &   $^{133}$I   &   20.8 h  &   C($\beta^-$) &   9.3(14) \\\hline
$^{97}$Zr   &   16.91 h &   C($\beta^-$) &   20(4)   &   $^{134}$I   &   52.6 m  &   C($\beta^-$) &   4.3(7)  \\\hline
$^{92m}$Nb  &   10.15 d &   I($\beta^+$,$\varepsilon$)  &   0.6(4)  &   $^{132}$Cs  &   5.48 d  &   I($\beta^{+/-}$)    &   6.2(17) \\\hline
$^{95}$Nb   &   34.98 d &   C($\beta^-$) &   17(2)   &   $^{136}$Cs  &   13.16 d &   C($\beta^-$) &   6.6(12) \\\hline
$^{96}$Nb   &   23.35 h &   I($\beta^-$) &   13.6(17)    &   $^{140}$Ba  &   12.75 d &   C($\beta^-$) &   3.6(6)  \\\hline
$^{97}$Nb   &   72.1 m  &   C($\beta^-$) &   13.3(21)    &   $^{140}$La  &   1.68 d  &   I($\beta^-$) &   7.4(12) \\\hline
$^{98m}$Nb  &   51.3 m  &   I($\beta^-$) &   13.9(19)    &   $^{135}$Ce  &   17.7 h  &   C($\beta^+$,$\varepsilon$)  &   17.0(25)    \\\hline
$^{99}$Mo   &   2.75 d &   C($\beta^-$) &   44(6)   &   $^{145}$Eu  &   5.93 d  &   C($\beta^+$,$\varepsilon$)  &   1.65(19)    \\\hline
$^{96}$Tc   &   4.98 d  &   C($\beta^+$,$\varepsilon$)  &   2.71(28)    &   $^{146}$Eu  &   4.59 d  &   C($\beta^+$,$\varepsilon$)  &   1.16(17)    \\\hline
$^{104}$Tc  &   18.3 m  &   C($\beta^-$) &   22(6)   &   $^{154}$Tb  &   21.5 h  &   C($\beta^+$,$\varepsilon$)  &   13(6)   \\\hline
$^{103}$Ru  &   39.26 d &   C($\beta^-$) &   63(10)  &   $^{156}$Tb  &   5.35 d  &   C($\varepsilon$)   &   1.7(5)  \\\hline
$^{105}$Ru  &   4.44 h  &   C($\beta^-$) &   34(5)   &   $^{198}$Au  &   2.70 d  &   C($\beta^-$) &   1.33(20)    \\\hline
$^{105}$Rh  &   35.36 h &   C($\beta^-$) &   77(13)  &   $^{206}$Bi  &   6.24 d  &   C($\varepsilon$)   &   1.21(25)    \\\hline
$^{106m}$Rh &   2.17 h   &   I($\beta^-$) &   15.0(23)    &   $^{240}$Am  &   50.8 h  &   I($\varepsilon$)   &   45(5)   \\\hline
\end{longtable}
}
\vskip -0.1cm
\noindent In the \textbf{Third Chapter} analysis of the experimental data using theoretical model calculations
is performed. In the analysis, calculations with the following computer models, based on the Monte-Carlo
method were employed~: \texttt{LAHET} (with Bertini and \texttt{ISABEL} cascades, Dresner evaporation and
Atchison fission (\texttt{RAL}), and also cascade of Cugnon \texttt{INCL} and fission by Schmidt \texttt{ABLA},
and \texttt{RAL}), Dubna cascade-evaporation-fission model \texttt{CASCADE} (by V.S. Barashenkov),
cascade-exciton models \texttt{CEM95} and \texttt{CEM2k} (S.G. Mashnik, LANL), as well as the four combined
models, obtained by joining the cascade parts of \texttt{CEM2k} or Los-Alamos version of quark-gluon model
\texttt{LAQGSM} (S.G. Mashnik, K.K. Gudima, LANL) with evaporation-fission parts of generalized evaporation
model by S.~Furihata \texttt{GEM2} and the binary fission model by R.~Charity \texttt{GEMINI}
(\texttt{LAQGSM+GEM2}, \texttt{CEM2k+GEM2}, \texttt{LAQGSM+GEMINI}, \texttt{CEM2k+GEMINI}).
Calculations using the programs not modelling fission products (\texttt{CEM95} and \texttt{CEM2k}) were
performed only for analysis of the data from the target ${}^{129}$I$ + {}^{127}$I. For analysis of
the cumulative cross sections, respective theoretical cross sections were also recalculated to cumulative
ones, based on known decay branching ratios. Calculations using the basic models of \texttt{LAHET} as well
as \texttt{CEM95} were performed by the author, while the other calculations were carried out by their
developers (group of V.S.~Barashenkov (JINR), group of S.G.~Mashnik, R.E.~Prael (LANL, USA)).

Qualitative by-isotope comparison of data from the $^{129}$I$ + {}^{127}$I with calculations using all
these models is shown in \textbf{Figure~\ref{figi}}. With the purpose of comparison of the experimental
and theoretical values a criterion proposed by R.~Michel as well as its standard deviation were used~:
$$
\left<F\right>=10^{\sqrt{\left<(\log\sigma^{exp}
-\log\sigma^{theo})^2\right>}}\,\,\,\,\mbox{and}\,\,\,\,
S(\left<F\right>)=10^{\sqrt{\left<\left(\left|\log\left(\frac{\sigma_{cal}^i}
{\sigma_{exp}^i}\right)\right|-\log(\left<F\right>)\right)^2\right>}}\,.
$$
Since a strong dependence of comparison results on the product mass number was observed, two mass regions
were chosen~: 1) all residuals with $A = 44 - 128$ and 2) only nuclei with $A \geqslant 95$
(see \textbf{Table~\ref{tableci}}). Out of 74 cross sections measured in this work, 42 were chosen for
such comparison [5,6,7,8] as those meeting the physics principles realized in the models.
Particularly, if only an isomer state of a nucleus was measured or, on the contrary, only its ground state
was measured whereas this nucleus possesses several rather long-lived isomers having significant branching
to the ground state, such cross sections were excluded from the comparison. If the cross sections of a
nucleus in both its metastable and ground state be experimentally determined independently, their sum
was compared with the calculated value.
\begin{table}[t]
\begin{center}
\caption{Theoretical model analysis of the residuals measured in
the target $^{127}$I$ + {}^{129}$I.}\label{tableci}
\begin{tabular}{|l|c|c|c|c|c|c|}  \hline
~~~Model  & \multicolumn{3}{|c|}{$A = 44 - 128$}
        & \multicolumn{3}{|c|}{$A \geqslant 95$} \\\cline{2-7}
        & $N/N_{30\%}/N_{2.0}$ & $\left <F\right>$ & $S(\left <F\right>)$
        & $N/N_{30\%}/N_{2.0}$ & $\left <F\right>$ & $S(\left <F\right>)$\\\hline
\texttt{LAHET} Bertini  & 36/6/22 &3.72  &  3.00 &   22/6/19& 1.67 &   1.34\\
\texttt{LAHET ISABEL}   & 34/5/18 &5.18  &  4.45 &   22/5/16& 1.72 &   1.37\\
\texttt{LAHET INCL+RAL} & 33/14/21&3.86  &  3.16 &   22/14/21&1.42 &   1.28\\
\texttt{LAHET INCL+ABLA}& 32/9/21 &9.32  &  7.01 &   22/9/21 &1.57 &   1.34\\
\texttt{CASCADE}        &42/9/15  &11.05 &  5.19 &   22/9/14 &3.32 &   2.75\\
\texttt{CEM95}          &40/10/20 &5.40  &  3.52 &   22/9/18 &1.78 &   1.44\\
\texttt{CEM2k}          &33/13/26 &2.89  &  2.74 &   22/11/20&1.48 &   1.27\\
\texttt{LAQGSM+GEM2}    &33/13/22 &3.16  &  2.68 &   22/13/21&1.50 &   1.34\\
\texttt{CEM2k+GEM2}     &35/10/28 &5.03  &  5.04 &   22/8/20 &1.60 &   1.35\\
\texttt{LAQGSM+GEMINI}  &42/19/29 &4.28  &  3.58 &   22/17/21&1.31 &   1.21\\
\texttt{CEM2k+GEMINI}   &42/12/27 &2.74  &  2.15 &   22/9/20 &1.46 &   1.25\\\hline
\end{tabular}
\end{center}
\end{table}
Results of the qualitative by-isotope analysis for the $^{237}$Np target are given in
\textbf{Figure~\ref{fignp}}.
For the quantitative comparison using the $\left <F\right>$ criterion, 37 out of 53
experimentally measured cross sections on $^{237}$Np were chosen. For the purpose of analysis,
the cross sections were divided into two groups: 1) belonging to the fission region
$48 \leqslant A \leqslant 175$, and 2) all the residuals [7,8] (see \textbf{Table~\ref{tablecnp}}).
\begin{table}[t]
\begin{center}
\caption{Theoretical model analysis of the residuals measured in
the target $^{237}$Np.}\label{tablecnp}
\begin{tabular}{|l|c|c|c|c|c|c|}  \hline
Модель  & \multicolumn{3}{|c|}{$A = 48 - 175$}
        & \multicolumn{3}{|c|}{$A = 48 - 234$} \\\cline{2-7}
        & $N/N_{30\%}/N_{2.0}$ & $\left <F\right>$ & $S(\left <F\right>)$
        & $N/N_{30\%}/N_{2.0}$ & $\left <F\right>$ & $S(\left <F\right>)$\\\hline
\texttt{LAHET} Bertini   &32/4/16 &3.29    &2.30    &37/6/20 &3.18    &2.26\\
\texttt{LAHET ISABEL}    &32/5/16 &3.34    &2.37    &37/5/19 &3.18    &2.26\\
\texttt{LAHET INCL+RAL}  &32/7/16 &3.51    &2.49    &37/8/19 &3.35    &2.39\\
\texttt{LAHET INCL+ABLA} &32/9/14 &4.09    &2.78    &37/9/14 &4.52    &2.92\\
\texttt{CASCADE}         &27/4/15 &9.75    &7.40    &32/5/16 &11.29   &7.75\\
\texttt{LAQGSM+GEM2}     &32/1/14 &7.28    &4.47    &37/2/15 &7.10    &4.29\\
\texttt{CEM2k+GEM2}      &32/5/16 &5.80    &3.70    &37/6/18 &5.43    &3.47\\
\texttt{LAQGSM+GEMINI}   &32/4/12 &4.69    &2.64    &36/5/13 &4.82    &2.69\\
\texttt{CEM2k+GEMINI}    &32/3/10 &3.57    &1.88    &36/4/11 &3.87    &2.05\\\hline
\end{tabular}
\end{center}
\end{table}
\begin{table}[b]
\begin{center}
\caption{Theoretical model analysis of the residuals measured in
the target $^{241}$Am.}\label{tablecam}
\begin{tabular}{|l|c|c|c|c|c|c|}  \hline
Модель  & \multicolumn{3}{|c|}{$A = 48 - 175$}
        & \multicolumn{3}{|c|}{$A = 48 - 240$} \\\cline{2-7}
        & $N/N_{30\%}/N_{2.0}$ & $\left <F\right>$ & $S(\left <F\right>)$
        & $N/N_{30\%}/N_{2.0}$ & $\left <F\right>$ & $S(\left <F\right>)$\\\hline
\texttt{LAHET} Bertini   &44/19/34    &2.28    &1.99    &37/6/20 &3.18    &2.26\\
\texttt{LAHET ISABEL}    &44/19/34    &2.30    &2.01    &37/5/19 &3.18    &2.26\\
\texttt{LAHET INCL+RAL}  &44/14/36    &2.44    &2.15    &37/8/19 &3.35    &2.39\\
\texttt{LAHET INCL+ABLA} &44/16/35    &2.96    &2.69    &37/9/14 &4.52    &2.92\\
\texttt{CASCADE}         &40/7/15     &6.57    &4.15    &32/5/16 &11.29   &7.75\\
\texttt{LAQGSM+GEM2}     &44/21/33    &3.38    &3.14    &37/2/15 &7.10    &4.29\\
\texttt{CEM2k+GEM2}      &44/21/32    &2.76    &2.50    &37/6/18 &5.43    &3.47\\
\texttt{LAQGSM+GEMINI}   &44/8/22     &3.69    &2.53    &36/5/13 &4.82    &2.69\\
\texttt{CEM2k+GEMINI}    &44/7/17     &3.18    &2.02    &36/4/11 &3.87    &2.05\\\hline
\end{tabular}
\end{center}
\end{table}
Results of the qualitative by-isotope analysis for the $^{241}$Am target are given in
\textbf{Figure~\ref{figam}}.
For the quantitative comparison using the $\left <F\right>$ criterion, 45 out of 80
experimentally measured cross sections on $^{237}$Np were chosen. For the purpose of analysis,
as before, the cross sections were divided into two groups:
1) belonging to the fission region
$48 \leqslant A \leqslant 175$, and 2) all the residuals [7,8]. The results of analysis
are given in \textbf{Table~\ref{tablecam}}.
\insnew{figi}{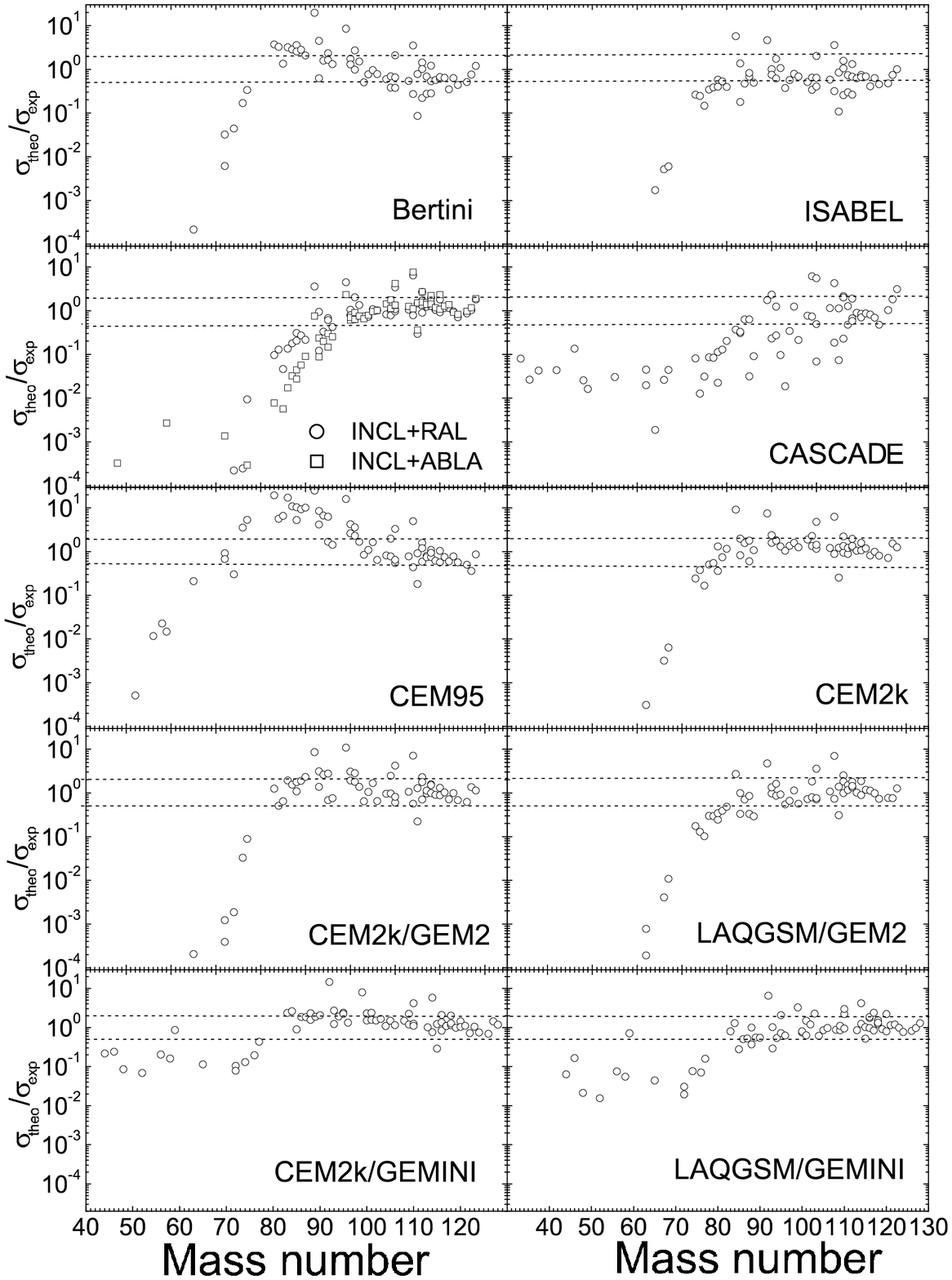}{Qualitative theoretical model analysis of the experimental residual
formation cross sections in the target 15\% $^{127}$I + 85\% $^{129}$I.}{0.0}{1.0}{1.2}

\insnew{fignp}{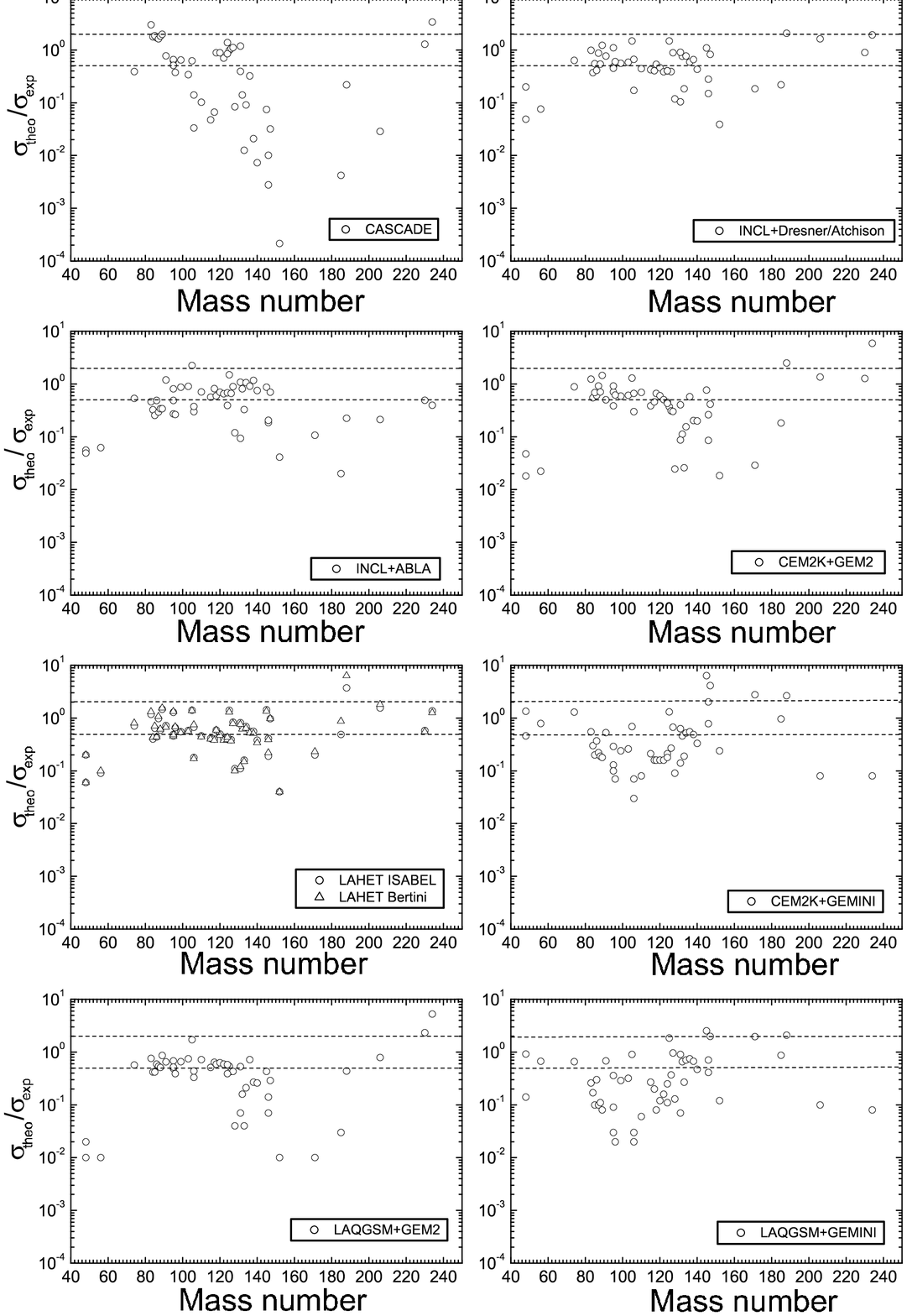}{Qualitative theoretical model analysis of the experimental residual
formation cross sections in the target $^{237}$Np.}{0.0}{1.0}{1.2}

\insnew{figam}{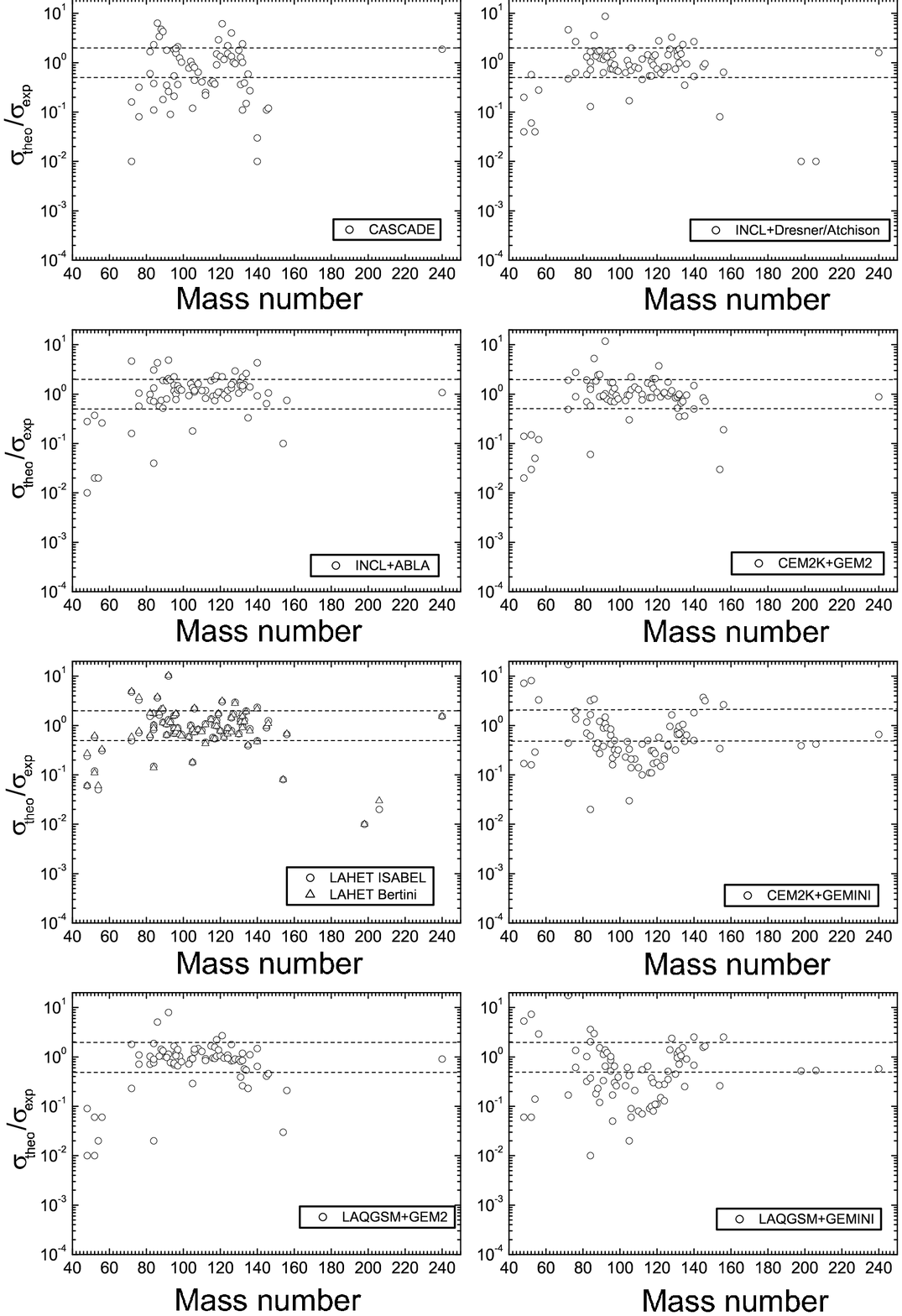}{Qualitative theoretical model analysis of the experimental residual
formation cross sections in the target $^{241}$Am.}{0.0}{1.0}{1.2}

\clearpage

\noindent In \textbf{Conclusion} the main results obtained in the work are summarized:
\begin{enumerate}
\item Developed and amended methods of short-lived $\beta$-unstable reaction product formation cross
section determination employing activation analysis with HPGe spectrometers, taking into account
a number of high-precision $\gamma$-spectroscopy approaches. Created and implemented a program package.
\item Carried out a high-precision measurement of $^{140}$La half-live with the use of a HPGe detector
under load change in a wide range, employing a procedure of deadtime underestimation correction.
Attained a half-live value in the echelon of accuracy of the recommended values.
\item Developed and demonstrated by the example of $A=152$ isobar a method for determination of
optimal experimental parameters for investigations of individual product nuclides formed with small
cross sections in hadron-nuclear interactions, also applicable for study of short-lived
$\beta$-unstable nuclei situated in complex decay chains.
\item Carried out experiments on determination by $\gamma$-spectroscopy formation cross sections
of product nuclei in interactions of 660-MeV protons with $^{129}$I, $^{237}$Np and $^{241}$Am
targets. Determined 74 residual nuclei in $^{129}$I target, 53 residual nuclei in $^{237}$Np, and
80 residual nuclei in $^{241}$Am. These data were obtained at intermediate energies for the
first time.

\item Carried out theoretical model analysis of the product nuclei formation cross sections in $^{129}$I,
$^{237}$Np, and $^{241}$Am targets using eleven existing models. Showed that for $^{129}$I all the
models generally give reasonable agreement with the data for the products with masses higher 95,
slightly better results being given by \texttt{CEM2k+GEMINI} and \texttt{LAQGSM+GEMINI} models (LANL, USA).
In the case of $^{237}$Np and $^{241}$Am targets, the best results were shown by Bertini and \texttt{ISABEL}
cascade models combined with evaporation-fission model \texttt{RAL}, all of them basic for \texttt{LAHET}
and \texttt{MCNPX} (LANL, USA) code systems. High precision of the data measured allowed to demonstrate
insufficient for practical applications accuracy of theoretical modelling of the reactions studied.
\end{enumerate}
\newpage
\noindent\textbf{List of Thesis publications}:
\def\refname{\vspace{-1em}}


\vspace{-1cm}
\begin{thebibliography}{99}
\bibitem{cite1} J. Adam, J. Mr\'azek, J. Fr\'ana, A. R. Balabekyan, V. S. Pronskikh,
V. G. Kalinnikov, A. N. Priemyshev, ``Ionizing Radiation Measurement: Software for Calculating Nuclear
Reaction Cross Sections'', Measurement Techniques \textbf{44}, 93 (2001).
\bibitem{cite2} J.~Adam, A.G.~Belov, R.~Brandt, P.~Chaloun, M.~Honusek, V.G.~Kalinnikov,
M.I.~Krivopustov, B.A.~Kulakov, E.-J.~Langrock, M.~Ochs, V.S.~Pronskikh, A.N.~Sosnin,
V.I.~Stegailov, V.M.~Tsoupko-Sitnikov, J.-S.~Wan, ``$^{140}$La half-life measurement
with Ge-detector'', Nuclear Instruments and Methods in Physics Research \textbf{B187}, 419-426 (2002).
\bibitem{cite3}  J.~Adam, A.~Balabekyan, V.S.~Pronskikh, V.G.~Kalinnikov, and  J.~Mr\'azek,
``Determination of the cross section for nuclear reactions
in complex nuclear decay chains'', Applied Radiation and Isotopes \textbf{56}, 607-613 (2002).
\bibitem{cite4} J.~Adam, A.~Balabekyan, R.~Brandt, V.P.~Dzhelepov, S.A.~Gustov, V.G.~Kalinnikov,
I.V.~Mirokhin, J. Mr\'azek, R.~Odoj, V.S.~Pronskikh, O.V.~Savchenko, A.N.~Sosnin,
A.A.~Solnyshkin, V.I.~Stegailov, V.M.~Tsoupko-Sitnikov, ``Investigation of the formation
of residual nuclei from the radioactive $^{237}$Np and $^{241}$Am targets in the
reaction with 660 MeV protons'', Physics of Atomic Nuclei \textbf{65}, 763-775 (2002).
\bibitem{cite5} V.S.~Pronskikh, J.~Adam, A.~Balabekyan, V.S.~Barashenkov, V.P.~Dzhelepov,
S.A.~Gustov, V.P.~Filinova, V.G.~Kalinnikov, M.I.~Krivopustov, I.V.~Mirokhin,
A.A.~Solnyshkin, V.I.~Stegailov, V.M.~Tsoupko-Sitnikov, J. Mr\'azek, R.~Brandt,
W.~Westmeier, R.~Odoj, S.G.~Mashnik, A.J.~Sierk, R.E.~Prael, K.K.~Gudima, M.I.~Baznat,
``Study of proton induced reactions in a radioactive $^{129}$I target at E$_p$=660 MeV'',
Proceedings of the International Workshop on Nuclear Data for the Transmutation of
Nuclear Waste (TRAMU@GSI), GSI-Darmstadt, Germany, September 1-5, 2003,
ISBN 3-00-012276-1, Eds: A. Kelic and K.-H. Schmidt; LANL Preprint LA-UR-04-2139, pp. 1-6,
E-print: nucl-ex/0403056, submitted to Physical Review C.
\bibitem{cite6} J.~Adam, V.S.~Barashenkov, V.P.~Dzhelepov, S.A.~Gustov, V.P.~Filinova, V.G.~Kalinnikov,
M.I.~Krivopustov, I.V.~Mirokhin, V.S.~Pronskikh, A.A.~Solnyshkin, V.I.~Stegailov, V.M.~Tsoupko-Sitnikov,
J. Mr\'azek, R.~Brandt, W.~Westmeier, R.~Odoj, S.G.~Mashnik, R.E.~Prael, K.K.~Gudima, M.I.~Baznat,
``Investigation of proton-nuclear reaction product formation in $^{129}$I target at 660-MeV proton energy''
(in Russian), ECHAYA Letters 4(121), 53-64 (2004).
\bibitem{cite7} J.~Adam, A.~Balabekyan, R.~Brandt, V.S.~Barashenkov, V.P.~Dzhelepov, V.P.~Filinova,
S.A.~Gustov, V.G.~Kalinnikov, M.I.~Krivopustov, I.V.~Mirokhin, J. Mr\'azek, R.~Odoj,
V.S.~Pronskikh, O.V.~Savchenko, A.N.~Sosnin, A.A.~Solnyshkin, V.I.~Stegailov,
V.M.~Tsoupko-Sitnikov, ``Investigation of the formation of residual nuclei in reactions
induced by 660 MeV protons interacting with the radioactive $^{237}$Np, $^{241}$Am and
$^{129}$I targets'', Journal of Nuclear Science and Technology, Supplement 2, 272-275 (August 2002).
\bibitem{cite8}  S.G.~Mashnik, V.S.~Pronskikh, J.~Adam, A.~Balabekyan, V.S.~Barashenkov,
V.P.~Filinova, A.A.~Solnyshkin, V.M.~Tsoupko-Sitnikov, R.~Brandt, R.~Odoj, A.J.~Sierk,
R.E.~Prael, K.K.~Gudima, M.I.~Baznat, ``Analysis of the JINR p(660 MeV) + $^{129}$I,
$^{237}$Np, and $^{241}$Am measurements with eleven different models'',
LANL Preprint LA-UR-04-4929, стр. 1-12, E-print: nucl-th/0407097,
Proceedings of the 7th Specialists' Meeting on Shielding Aspect of Accelerators,
Targets and Irradiated Facilities, SATIF-7, Sacavem (Lisbon), Portugal, May 17-18, 2004.
\end{thebibliography}
\end{document}